# Determination of the scaling characteristics of time-dependent optical properties of microalgae using electromagnetic scattering

CHUNYANG MA[1,*]

[1]*School of Mechanical and Electrical Engineering, Nanchang University, Nanchang 330031, P. R. China*
*Corresponding author: cyma@ncu.edu.cn*

**The time-dependent optical properties of microalgae are crucial for light transfer in photobioreactor (PBR) designs. Here, the time-dependent optical properties were derived using electromagnetic scattering theory informed by the experimentally measured optical properties of *Chlorella protothecoides*. The temporal scaling functions (TSFs) of nearly wavelength-independent absorption and scattering cross-sections were demonstrated using the electromagnetic scattering theory, leading to the first concrete expression of the TSF. The TSF establishes the relationship between the time-dependent absorption/scattering cross-sections at the stationery and growth phases of microalgal development. The concrete expression of the TSF provides a new way to calculate the time-dependent optical properties of microalgae using electromagnetic scattering theory. The areas where the TSF of microalgae has exciting potential are in remote sensing and photobioreactor applications.**

Microalga cultivation can produce medicines, health drinks, animal feed, and other value-added products [1,2]. Moreover, it represents a sustainable energy source with the potential to help alleviate climate change by providing an alternative to energy sources that generate high $CO_2$ emissions [3,4]. The optical properties of microalgae are important parameters that are crucial to their light transfer abilities, photosynthetic efficiency, and electromagnetic scattering behavior.

The light utilization efficiency of microalgae in photobioreactors (PBRs) affects the biofuel production process significantly [5–7]. An optimum irradiance that achieves maximum photosynthesis efficiency in microalga cultivation has been reported [7]. However, the light irradiance in PBRs depends on the time-dependent optical properties of microalgae, implying that the optimum irradiance varies during the growth process of microalgae. Early research focused on measuring the optical properties of microalgae at the stationary stage [8–10]. Recently, the time-dependent optical properties of microalgae have been measured in batch cultures [11–13]. A theory for the temporal scaling of growth-dependent optical properties in microalgae was proposed by Zhao et al., which provides a mathematical proof of temporal scaling characteristics, but does not give the specific form and determination method of temporal scaling functions [13]. Nevertheless, a time-dependent light transfer analysis is required in order to design and optimize light distribution in PBRs, and thus improve photosynthetic efficiency.

The aim of this study is to demonstrate a temporal scaling function (TSF) of time-dependent absorption and scattering cross-sections that is largely independent of wavelength using electromagnetic theory. The derived mathematical expression provides a deep insight into the temporal scaling characteristics of the time-dependent optical properties and the origin of electromagnetic scattering in microalgae. Furthermore, the complex refractive index was retrieved from the measured time-dependent absorption and scattering cross-sections and cell size distribution data for *Chlorella protothecoides*. The results can help to calculate the time-dependent optical or optical properties of multiple microalgae species.

Light transfer within absorbing, scattering, and non-emitting microalgal suspension in photobioreactors is governed by the radiative transfer equation [14]:

$$\mathbf{s} \cdot \nabla I(\mathbf{r},\mathbf{s}) + \beta I(\mathbf{r},\mathbf{s}) = \frac{\kappa_s}{4\pi} \int_{\Omega'=4\pi} I(\mathbf{r},\mathbf{s}) \Phi(\mathbf{s}' \to \mathbf{s}) \mathrm{d}\Omega' \quad (1)$$

where $I$ is the radiative intensity in the direction $\mathbf{s}$ at location $\mathbf{r}$, $\mathbf{s}$ is the direction vector, $\kappa_s$ is the scattering coefficient, $\beta = \kappa_a + \kappa_s$ is the extinction coefficient, $\kappa_a$ is the absorption coefficient, $\Phi(\mathbf{s}',\mathbf{s})$ is the scattering phase function, and $\Omega'$ is the solid angle. The scattering phase function $\Phi(\mathbf{s}',\mathbf{s})$ represents the probability that the radiation transfer in the solid angle $\mathrm{d}\Omega'$ around the direction $\mathbf{s}'$ is scattered into the solid angle dΩ around the direction **s**. The average $\langle C_{abs} \rangle$ absorption and $\langle C_{sca} \rangle$ scattering cross-sections of a polydisperse microalgal suspension are related to the absorption and scattering cross-sections, respectively, by the cell size distribution function, and are expressed as [14]

$$\langle C_{abs} \rangle = \int C_{abs}(d_s) f(d_s) \, d d_s \quad (2a)$$

$$\langle C_{sca} \rangle = \int C_{sca}(d_s) f(d_s) \, d d_s \quad (2b)$$

where $C_{abs}(d_s)$ and $C_{sca}(d_s)$ are the absorption and scattering cross-sections of a single spherical scatterer with diameter $d_s$, respectively, and $f(d_s)$ is the cell size distribution function.

The microalgae are assumed to have an axisymmetric spheroidal shape with major and minor diameters $a$ and $b$, respectively, which are approximated as spheres with an equivalent diameter $d_s$, so that their surface area is identical to that of their actual spheroidal shape. The equivalent diameter of the sphere with the same surface area as the spheroid is expressed as [15]

$$d_s = \frac{1}{2}\left(2a^2 + 2ab \frac{\sin^{-1} e}{e}\right)^{1/2} \quad \text{where} \quad e = \frac{(\varepsilon^2 - 1)^{1/2}}{\varepsilon} \quad (3)$$

Here, $\varepsilon$ is the spheroid aspect ratio, defined as $\varepsilon = a/b$. Bidigare et al. and Pottier et al. used a predictive method to determine the spectral absorption coefficient $\kappa_{a,\lambda}$ by [16, 17]

$$\kappa_{a,\lambda} = \sum_{i=1}^{n} Ea_{\lambda,i} c_i \quad (4)$$

where $Ea_{\lambda,i}$ (m²/kg) is the *in vivo* pigment-specific spectral absorption cross-section of pigment $i$, and $(c_i)_{1 \le i \le n}$ are the mass concentrations (kg/m³) of the cell's pigments. The specific absorption cross-sections of chlorophyll (Chl) a, b, c, and $\beta$-carotene have been reported by Bidigare et al. in the 400–750 nm spectral region [18].

The radiation characteristics of microalgae are related to cell morphology, size, internal structure, components, and cell size distribution [19]. During microalgal growth a number of processes occur, including the division of cells, changes in pigments [11], and variations in levels of carbohydrates, proteins, and lipids [20]. Hence, the optical properties of microalgae vary during the growth period. The complex refractive index $m = n + k\mathrm{i}$ is close to unity in water because the cells in living microalgae are composed mostly of water. Following the approach of van de Hulst [21], the anomalous diffraction approximation theory can be used to obtain the absorption efficiency factor in this special condition :

$$Q_{abs}(\rho) = 1 + [\exp(-2\rho \tan \xi)(2\rho \tan \xi + 1) - 1]/2\rho^2 \tan^2 \xi \quad (5)$$

where $Q_{abs}(\rho)$ represents the absorption efficiency factor, $\rho = 2\alpha(n-1)$, $\tan \xi = k/(n-1)$, and $\alpha = \pi d n_w/\lambda$, where $\lambda$ is the wavelength in a vacuum. By introducing the parameter $\rho' = 4\alpha k = 2\rho \tan \xi$, the absorption efficiency factor can also be rewritten as [21]

$$Q_{abs}(\rho') = 1 + 2 \frac{\exp(-\rho')}{\rho'} + 2 \frac{\exp(-\rho') - 1}{\rho'^2} \quad (6)$$

When $\rho'$ is much smaller than 1, we can use the Taylor series expansion for $Q_{abs}(\rho')$ to obtain the following result:

$$Q_{abs}(\rho') = \frac{2}{3}\rho' + o(\rho'^2) \quad (7)$$

Considering the temporal evolution of the optical properties of microalgae, $\rho' = 4\alpha k$ represents the growth time function $t$; hence, the temporal scaling factor of absorption can be expressed as

$$R_{abs}(t_a, t_s) = \frac{k(t_a) \int d^3(t_a) f(d(t_a)) \, d d}{k(t_s) \int d^3(t_s) f(d(t_s)) \, d d} \quad (8)$$

where the absorption index $k$ is a function of wavelength $\lambda$. According to Eq. (4), the ratio of $k(t_a)/k(t_s)$ can be expressed as

$$\frac{k(t_a)}{k(t_s)} = \frac{\kappa_{a,\lambda}(t_a)}{\kappa_{a,\lambda}(t_s)} = \frac{\sum_{i=1}^{n} Ea_{\lambda,i} c_i(t_a)}{\sum_{i=1}^{n} Ea_{\lambda,i} c_i(t_s)} \quad (9)$$

using the relation $\kappa_{a,\lambda}(t) = 4\pi k(t)/\lambda$ [22] because the microalgal cells are highly forward scattering [9]. In the case that the mass concentrations of the cell's pigments increase or decrease consistently with time, that is, satisfying the relation $c_i(t_a) = K(t) c_i(t_s)$ for $1 \le i \le n$, the ratio of $k(t_a)/k(t_s)$ is independent of wavelength $\lambda$. Furthermore, the temporal scaling factor of absorption $R_{abs}(t_a, t_s)$ is independent of wavelength $\lambda$ over the Vis-NIR spectral region.

The analysis of $Q_{sca}(\rho)$ utilizes the transport approximation proposed by Dombrovsky et al. for large semi-transparent particles :

$$Q_{sca}^{tr}(\rho) = C \begin{cases} \rho/5 & (\rho \le 5) \\ (5/\rho)^\gamma & (\rho > 5) \end{cases} \quad (10)$$

where $C = 1.5n(n-1)\exp(-15k)$ and $\gamma = 1.4 - \exp(-80k)$. For microalgal cells, the value of $\rho = 2\alpha(n-1)$ is usually smaller than 5. Therefore, the following equation can be used [23]:

$$Q_{sca}^{tr}(\rho) = 1.5n(n-1)\exp(-15k)(\rho/5) \quad (11)$$

$$Q_{sca}(\rho) = \frac{Q_{sca}^{tr}(\rho)}{1 - g} \quad (12)$$

Considering the temporal evolution of the optical properties of microalgae, $\rho = 2\alpha(n-1)$ is a function of growth time $t$; therefore, the temporal scaling factor of scattering can be expressed as

$$R_{sca}(t_a, t_s) = \frac{\frac{\pi}{4} \int Q_{sca}(\rho(t_a)) d^2(t_a) f(d(t_a)) \, d d}{\frac{\pi}{4} \int Q_{sca}(\rho(t_s)) d^2(t_s) f(d(t_s)) \, d d} \quad (13)$$

Inserting the expression for $Q_{sca}(\rho)$ leads to $R_{sca}(t_a, t_s)$ being rewritten as

$$R_{sca}(t_a,t_s) = \frac{n(t_a)(n(t_a)-1)^2 \int d^3(t_a) f(d(t_a))\,\mathrm{d}d}{n(t_s)(n(t_s)-1)^2 \int d^3(t_s) f(d(t_s))\,\mathrm{d}d} \quad (14)$$
$$\times \exp\{15[k(t_s)-k(t_a)]\}$$

Here, the term $\exp\{15[k(t_s)-k(t_a)]\} \approx 1$, as $15[\dot{k}(t_a)(t_s-t_a)] \ll 1$ for microalga cultivation. Alternatively, we can choose a time interval infinitely close to 0. In this event, $R_{sca}(t_a,t_s)$ can be rewritten as

$$R_{sca}(t_a,t_s) = \frac{n(t_a)(n(t_a)-1)^2 \int d^3(t_a) f(d(t_a))\,\mathrm{d}d}{n(t_s)(n(t_s)-1)^2 \int d^3(t_s) f(d(t_s))\,\mathrm{d}d} \quad (15)$$

Assuming that the refractive index $n$ satisfies some proportional relation, such as the absorption index, then the temporal scaling factor of scattering $R_{sca}(t_a,t_s)$ will be independent of wavelength $\lambda$ over the Vis-NIR spectral region. In addition, it is noted that the refractive index remains approximately constant over the Vis-NIR region [24]. Kandilian et al. [10] verified that the scattering phase function demonstrated near-independence with respect to wavelength over the photosynthetically active region. Ma et al. [25] reported that the scattering phase function is nearly time-independent during culture cultivation. Thus, the scattering phase function can be considered independent of the wavelength and growth time.

In this study, microalgae were considered to be approximately spherical in shape. Therefore, we used the Lorentz–Mie theory to calculate the radiation characteristics of the microalgae cell. Fig. 1 shows a schematic diagram of the inverse procedure of the optical constants. The Lorentz–Mie theory [22] was employed in the forward model to calculate the average extinction and absorption cross-sections. The trust region algorithm was used to minimize the difference between the simulated and measured absorption and extinction cross-sections of the microalgae. For each wavelength, the objective function $\eta_\lambda$ is defined as

$$\eta_\lambda = \left(\frac{\langle C_{abs,\lambda,pred}\rangle - \langle C_{abs,\lambda,exp}\rangle}{\langle C_{abs,\lambda,exp}\rangle}\right)^2 + \left(\frac{\langle C_{ext,\lambda,pred}\rangle - \langle C_{ext,\lambda,exp}\rangle}{\langle C_{ext,\lambda,exp}\rangle}\right)^2 \quad (16)$$

The convergence criteria were set as $\eta_\lambda < 10^{-4}$. To verify the theoretical approach, the complex refractive index of *Chlorella sp.* was retrieved using the experimentally measured cell size distribution and the absorption and extinction cross-sections reported in [8]. Comparing (not shown) the measured complex refractive index of *Chlorella sp.* with that reported by Lee et al. [15] indicates that the relative errors associated with the real and imaginary parts are less than 0.43% and 33.14% (absolute error less than 5×10$^{-4}$), respectively, for the 400–750 nm wavelength range, thus confirming the validity of the inverse model; this model assumes that the microalga solution has a refractive index equal to the distilled water in the 380–850 nm spectral region.

Figure 2a shows the wavelength dependence of the complex refraction index of microalgae at different growth stages, which was calculated using experimentally measured major and minor diameters, via Eq. (3), as well as the absorption and scattering cross-sections [13]. The complex refraction index varies significantly with the growth time, reflecting the change in microalga composition at different growth stages. The absorption index, shown in Fig. 2b, displays peaks at approximately 435 and 676 nm corresponding to absorption peaks of Chl a, and at 485 nm, which is indicative of carotenoids [11]. Interestingly, dips in the refractive index can be observed around the wavelengths corresponding to the absorption peaks; this is attributed to the fact that the real and imaginary parts of the complex refractive index are not independent of each other [19].

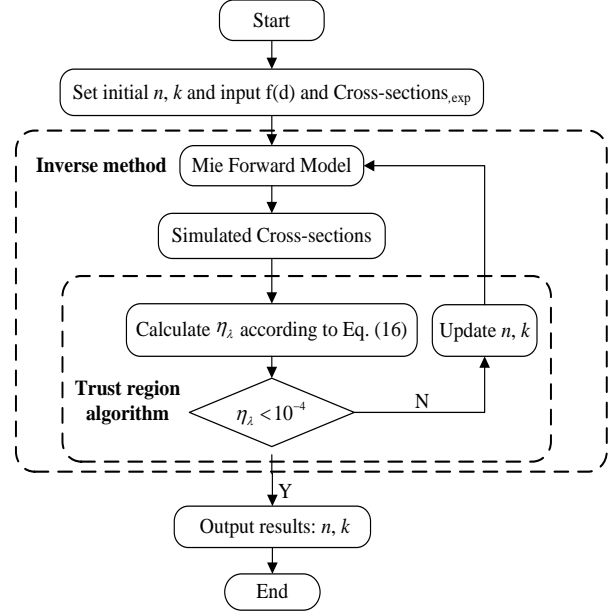

Fig. 1. Block diagram of the procedure used to retrieve the complex refractive indices of microalgae from cell size distribution data and the absorption/extinction cross-sections.

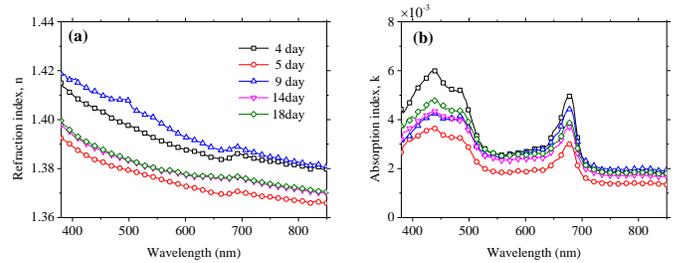

Fig. 2. The time-dependent refraction and absorption indices of *C. protothecoides* over the spectral region 380 to 850 nm.

Zhao et al. [13] reported that a connection exists between the time-dependent optical properties of microalgae, which is called the temporal scaling function. Here, the temporal scaling characteristics of absorption and extinction are largely invariant over the Vis-NIR spectral region, as demonstrated using electromagnetic scattering theory. Figure 3 presents the temporal scaling factors of absorption (Fig. 3a) and scattering (Fig. 3b) predicted using Eq. (8) and Eq. (15), respectively. As shown, these temporal scaling factors are nearly independent of the wavelength at 5 and 14 days in the growth process. Conversely, wavelength-independent behavior is not obviously observed at 4 and 9 days, which is likely due to the variation of the components and pigments present in the microalgae at these growth stages [11]. It can be inferred that the optical properties of microalgae vary faster than the experimentally measured time interval. The growth process of microalgae is extremely complex, and this study seeks to propose a method to calculate time-dependent optical properties.

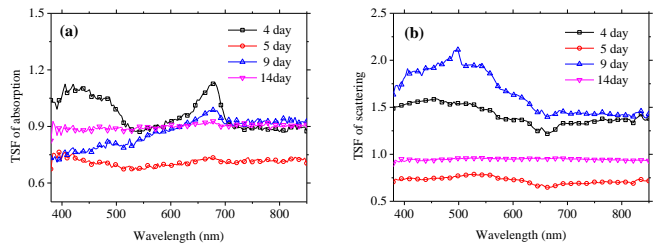

Fig. 3. The temporal scaling factors of absorption (a) and scattering (b) predicted under electromagnetic scattering.

The time-dependent optical properties of *C. protothecoides* batch cultures are presented in the 380–850 nm spectral range. These time-dependent optical properties are shown to vary significantly with growth time. The evolution of the time-dependent optical properties is attributed to changes in cell composition. To elucidate the evolution of microalgal optical properties, the temporal scaling characteristics between the growth phase and that at the stationary phase were established using electromagnetic scattering theory. The temporal scaling characteristics discussed in this study are likely generally applicable to microalgae owing to the fact that most strains possess a complex refractive index that is near unity in water. Moreover, the temporal scaling characteristics of microalgae may have important applications in remote sensing and photobioreactor design. Future investigations should focus on isolating the specific impacts that various microalgal components, e.g., pigments, proteins, lipids, etc., have on the complex refractive index of microalgal colonies.

**Funding sources and acknowledgments.** Nanchang University, China

**Acknowledgments.** I would like to acknowledge the support provided by Nanchang University and the support of my family.